\documentclass[twocolumn]{biophys}
\usepackage{helvet,times}
\usepackage{bm,textcomp}
\usepackage{amsmath}
\usepackage[round,numbers,sort&compress]{natbib}

\gridframe{N}
\cropmark{N}
\newcommand{\kb}{k_\text{\tiny B}}

\newcommand{\g}{\gamma}
\newcommand{\ts}{\tau_\text{c}}

\newcommand{\Da}{D_\text{\tiny A}}
\newcommand{\A}{\text{\tiny A}}

\providecommand{\avg}[1]{\left \langle #1 \right \rangle}

\begin{document}

\title{Spatial Fluctuations at Vertices of Epithelial Layers:\\Quantification of Regulation by Rho Pathway}

\author{\'E. Fodor,$^{1,2,*}$ V. Mehandia,$^{3,4,5,*}$ J. Comelles,$^{3,4}$ R. Thiagarajan,$^{3,4}$ N. S. Gov$^6$\\P. Visco,$^2$ F. van Wijland,$^2$ D. Riveline$^{3,4}$}

\address{\small{
$^1$DAMTP, Centre for Mathematical Sciences, University of Cambridge, Cambridge, United Kingdom; $^2$Laboratoire Mati\`ere et Syst\`emes UMR 7057 CNRS/P7, Universit\'e Paris Diderot, Paris cedex 13, France; $^3$Laboratory of Cell Physics, ISIS/IGBMC, Universit\'e de Strasbourg and CNRS (UMR 7006), Strasbourg, France; $^4$Development and Stem Cells Program, IGBMC, CNRS (UMR 7104), INSERM (U964), Universit\'e de Strasbourg, Illkirch, France; $^5$School of Mechanical, Materials and Energy Engineering, Indian Institute of Technology, Ropar, India; and $^6$Department of Chemical Physics, Weizmann Institute of Science, Rehovot, Israel}
}

\begin{abstract}

{In living matter, shape fluctuations induced by acto-myosin are usually studied {\it in vitro} via reconstituted gels, whose properties are controlled by changing the concentrations of actin, myosin and cross-linkers. Such an approach deliberately avoids to consider the complexity of biochemical signaling inherent to living systems. Acto-myosin activity inside living cells is mainly regulated by the Rho signaling pathway which is composed of multiple layers of coupled activators and inhibitors. We investigate how such a pathway controls the dynamics of confluent epithelial tissues by tracking the displacements of the junction points between cells. Using a phenomenological model to analyze the vertex fluctuations, we rationalize the effects of different Rho signaling targets on the emergent tissue activity by quantifying the effective diffusion coefficient, the persistence time and persistence length of the fluctuations. Our results reveal an unanticipated correlation between layers of activation/inhibition and spatial fluctuations within tissues. Overall, this work connects the regulation via biochemical signaling with mesoscopic spatial fluctuations, with potential application to the study of structural rearrangements in epithelial tissues.}

{$^*$These authors contributed equally to this work.}

\end{abstract}

\maketitle


\section*{INTRODUCTION}

Changes in shapes of cells and tissues are mediated by the acto-myosin cytoskeleton. To reproduce the dynamics of this network, minimal systems made of actin filaments, myosin motors and cross-linkers are synthetized {\it in vitro}~\cite{Backouche2006, Betz:14, Blanchoin2014, Murrell2015, Krauger2016}. The mechanics and dynamics of such {\it active gels} are controlled by varying the concentration of their various components. Activity of each component is monitored by adding some inhibitor drugs, and/or by tuning the ATP concentration of the system. Recent experimental evidence have shown the relevance of this approach to investigate the role of motors and cross-linkers in the emerging properties of the network~\cite{Silva:11, Betz:14}. These studies are based on tracking the motion of tracers injected in active gels: analyzing the spontaneous fluctuations of such tracers enables one to extract information about the activity of internal motors.

In multicellular systems, such as tissues of developing embryos, acto-myosin drives the {\it morphogenesis}: dramatic rearrangements leading to the formation of distinct organs~\cite{Rauzi:08, Guillot:2013}. This remodelling is mainly under the control of intracellular activity, which powers spatial fluctuations~\cite{Marmottant}, and intercellular interactions mediated by adhesion between neighboring cells~\cite{Cavey, Bardet:2013}. In contrast to synthetic gels, the internal regulation of the cellular acto-myosin activity is more complex {\it in vivo}. Therefore, extending the {\it in vitro} approach, based on controlling externally the activity of each specific component, to {\it in vivo} situations requires new strategies.

The Rho signaling pathway is known to regulate the acto-myosin activity in living cells~\cite{Hall2002}. It also controls cell-cell junctions~\cite{Reyes2014} and the elasticity of stress fibers~\cite{Oakes2016}. Such a pathway can be viewed as a series of activators and inhibitors installing a hierarchy of potential targets~\cite{Gibson:06, Riveline}. Activations and inhibitions controlled by each target are such that anticipating their net effects on the tissue fluctuations, powered by acto-myosin activity, remains a challenge~\cite{Besser2007, Bement2015}. In that respect, the inherent complexity of internal activity {\it in vivo} calls for new experiments and quantitative analysis to bridge the biochemical signaling of the Rho pathway with the emerging tissue dynamics.

In this paper, we explore the regulation of active fluctuations by the Rho pathway in epithelial monolayers. We measure these fluctuations by tracking tricellular junctions or vertices over time. In contrast with active gels, our analysis of internal fluctuations does not require to inject external tracers. Based on a phenomenological model, we quantify key parameters of junction activity: their effective diffusion coefficient, as well as the persistence time and persistence length of spatial fluctuations. We report modifications of these parameters for various targets along the signaling pathway. These results support that, for the inhibitions that we considered, the active fluctuations of the vertices are reduced when going downstream in the Rho pathway inhibition.


\section*{MATERIALS AND METHODS}

Experiments were performed with MDCK II cells stably expressing E-cadherin GFP (Nelson Lab.). We culture cells in DMEM containing $10\%$ Fetal Calf Serum (FCS) and antibiotics. We replate them on glass coverslips (CS) of $25$~mm diameter for live cell imaging. When the cell monolayer covered $70\%$ of the CS area, we firmly place the sample at the bottom of a custom made metallic holder. For acquisition, we change the medium to L15, $10\%$ FCS and antibiotics. We use the following inhibitors from myosin up to Rho at optimal concentrations following the manufacturer recommendations: inhibition of acto-myosin by ML-7 (Sigma-Aldrich, $10$~$\mu$M), inhibition of Rho kinase (ROCK) by Y-27632 (Sigma-Aldrich, $10$~$\mu$M), and inhibition of Rho by C3 Transferase (Cytoskeleton, $0.04$~$\mu$M). Note that the use of blebbistatin to inhibit directly myosins yielded some detrimental effects, leading us to rather use ML-7 instead.

For observation, we use a motorized inverted microscope (Nikon Eclipse Ti), equipped with a $12$~bit CCD camera (Photometric CoolSNAP HQ2). The setup is temperature controlled at $37^\circ$C (Life Imaging Services). We check with fluorescent beads ($4$~$\mu$m, TetraSpeck, Invitrogen Molecular Probes) grafted on CS surface that no drift appears during $24$~hours of live imaging after $2$~hour stabilization. We take pictures of the monolayer every $5$~min during the next $8$~hours with multiple z-stacks $1$~$\mu$m apart. They span $3$~$\mu$m depth of the cell monolayer. We merge the z-stacks into one image by using the maximum intensity projection. We then extract vertex positions from the sequence of merged images by manually clicking in each frame as long as they are visible. The procedure was validated in its precision through automatised detection as well. For each condition, we check that the average cell area was always about $180\pm15$~$\mu$m$^2$, and we consider more than $20$ vertices for at least $3$ biological repeats.


\section*{RESULTS}

\section*{Vertex tracking and inhibitors in the Rho pathway}

We use Madin Darby Canine Kidney (MDCK) cells stably transfected with E-cadherin fused with the Green Fluorescent Protein (GFP) as a paradigm for epithelial tissues dynamics~\cite{Adams}. This allows us to study live cells while interacting with each other. We seek to identify spatial points primarily involved in tissue transformations. The meeting points between three cells are involved in exchanges between neighbouring cells, thus serving as hallmark of tissue dynamics~\cite{Bardet:2013, Salomon2017}. Some specific proteins, such as tri-cellulin, are known to accumulate at this point in cell culture. Besides, vertices have also attracted the attention of developmental biologists which exhibit accumulation of proteins at these specific points~\cite{Lye2014}.

\begin{figure}[b]
\includegraphics[width=\columnwidth]{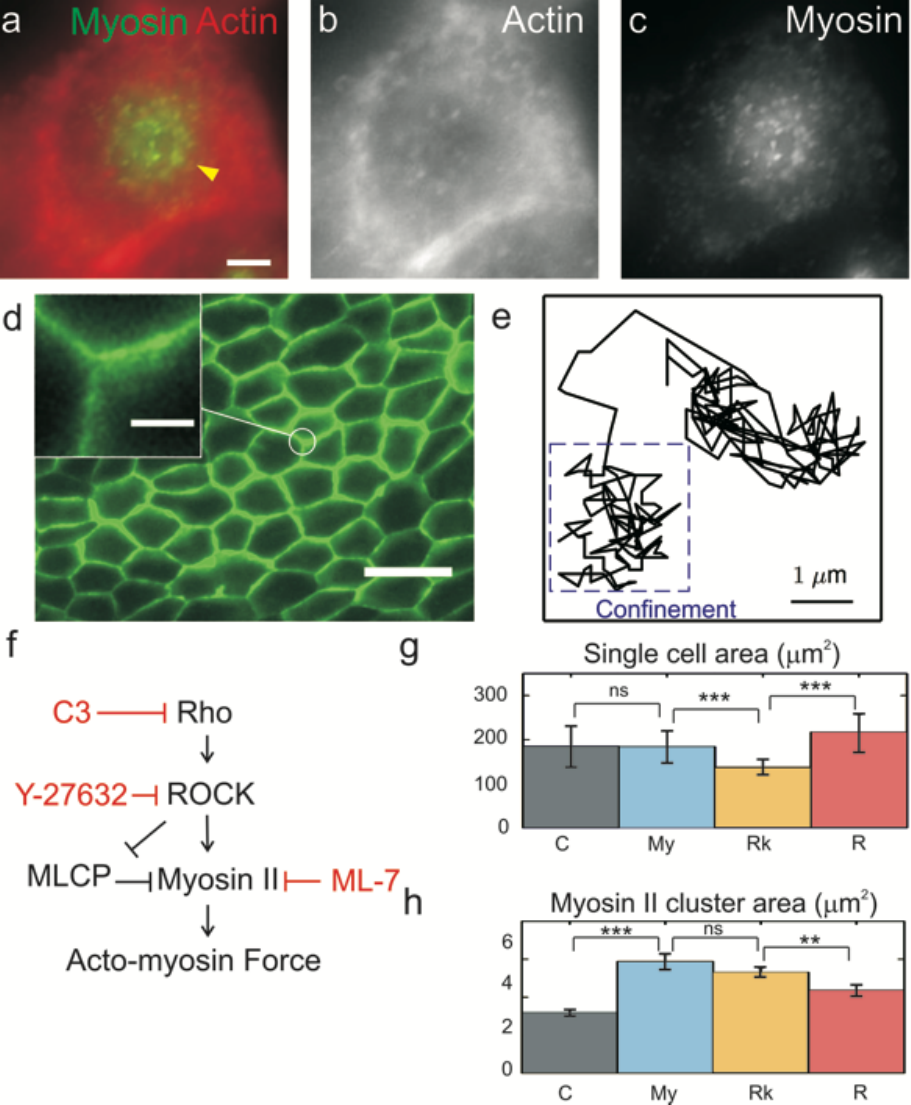}
\caption{\label{fig:exp}
	Study of vertex fluctuations.
	(a)~Actin (red) and myosin (green) structures at the apical surface of a MDCK cell (scale bar 3 $\mu$m). The myosin is concentrated in dense contractile units (yellow arrow) referred to as myosin clusters.
	(b)~Actin structure alone.
	(c)~Myosin structure alone.
	(d)~We visualise MDCK cell monolayer by GFP E-cadherin (scale bar $30$~$\mu$m).
	(Inset)~We identify the meeting points between three cells as the privileged point for our analysis (scale bar $4$~$\mu$m).
	(e)~Extraction of a typical transition between two locally stable positions in vertex trajectory (total time 8 hours).
	(f)~Simplified diagram of the Rho pathway installing an order relation in myosin activation, as presented in~\cite{Riveline}; in red the specific inhibitors and their targets.
	(g)~Area of individual cells in each condition. C: Control; My: Myosin inhibitor; Rk: Rho kinase (ROCK) inhibitor; R: Rho inhibitor. Number of experiments $\times$ number of cells = C: $4\times235$; My: $4\times246$; Rk: $3\times249$; R: $3\times148$.
	(h)~Area of myosin clusters. Number of experiments $\times$ number of clusters = C: $2\times312$; My: $2\times186$; Rk: $2\times197$; R: $2\times257$.
	Statistical analysis with one-way ANOVA test:  ns (non significant) $p > 0.05$, * $p < 0.05$, ** $p < 0.01$, *** $p < 0.001$ (see the Supporting Material).
}
\end{figure}

Vertex dynamics are driven both by thermal fluctuations and by active fluctuations powered by some internal nonequilibrium processes such as motor-induced forces, actin polymerization and cell-cell adhesion. The myosin-II motors are localized in dense contractile units present in the apical surface of the tissue~\cite{Rauzi:2010, Klinger:2014} [Figs.~\ref{fig:exp}(a-c)]. The active forces lead to large displacements of the vertices distinct from the thermal fluctuations of smaller amplitude. We focus here on large displacements which do not lead to any topological transitions in the tissue [Figs.~\ref{fig:exp}(d-e)]. The Rho signaling pathway controls the acto-myosin activity inside cells, namely the forces induced by myosin and/or by actin polymerization (see Fig.~9 in~\cite{Riveline}). In this study, we focus on the downstream targets that affect myosin to establish the principles for the validity and relevance of our framework. Upstream and downstream targets install a hierarchy in the activation of myosin.  We specifically inhibit the following targets: Rho, Rho kinase (ROCK) and myosin-II [Fig.~\ref{fig:exp}(f)]. Each inhibitor is specific to its target and incubated at the optimal concentrations for its inhibition. We used standard concentration values already utilized in other cell biology studies for many cell types including MDCK~\cite{Jin2001, Silberzan2014, Muller2015, Imai2015}. Altogether, we probe four conditions on the same system by considering untreated cells as a control.

To demonstrate that the chosen inhibitors and their concentration specifically act on the contractile state of cells, we investigate their effect on both architecture and contractile forces within tissues by measuring single cell area and myosin cluster area in each condition. The distribution of single cell area is not strongly affected in the myosin-II inhibitor case compared with control, as apparent from the mean value in Fig.~\ref{fig:exp}(g). The distribution gets modified in the Rho kinase and Rho inhibitor cases, with mean values being slightly reduced and increased, respectively. In contrast, the polygonicity distribution remains approximately the same for all conditions at different times, supporting that tissue architecture is barely affected by external inhibitors (see Fig.~S1 in the Supporting Material).

Myosin clusters have been shown to be relevant read-outs to assess the contractile state of cells~\cite{Munjal2014, Wollrab2016}. Along this line, measurements of cluster characteristics are informative. In our case, the mean density value of myosin clusters for control is larger than in the inhibited cases (see Fig.~S2 in the Supporting Material), suggesting a decrease in force generation for the same level of myosin per cell. In addition, the area of each myosin cluster is smaller in control than in other conditions [Fig.~\ref{fig:exp}(h)]. Besides, this area increases the more downstream along the Rho pathway inhibition, suggesting a relaxation of the myosin pool in the apical side, consistently with the notion that myosin-induced forces are reduced. Altogether, our analysis of clusters confirms that we are acting on the contractile state of cells. Since inhibitors are specific and used at their optimal concentrations, these measurements support the validity of our experimental approach.


\section*{Statistics of vertex displacement: inhibitors affect spatial fluctuations}

Our goal is to investigate how the emergent fluctuations of the tissue are regulated by the Rho pathway. To this aim, we first demonstrate that the inhibitors in the pathway affect these fluctuations by extracting the statistics of displacements from the vertex trajectories. This allows us to assess the existence of a direct link between biochemical signaling and mechanical fluctuations. In our analysis, we do not consider neither spatial inhomogeneities nor topological transitions that occur in the tissues. In that respect, we measure vertex trajectories in the absence of neighboring cell division, by tracking them as long as they are visible until a maximum of 8 hours.

\begin{figure}
\includegraphics[width=\columnwidth]{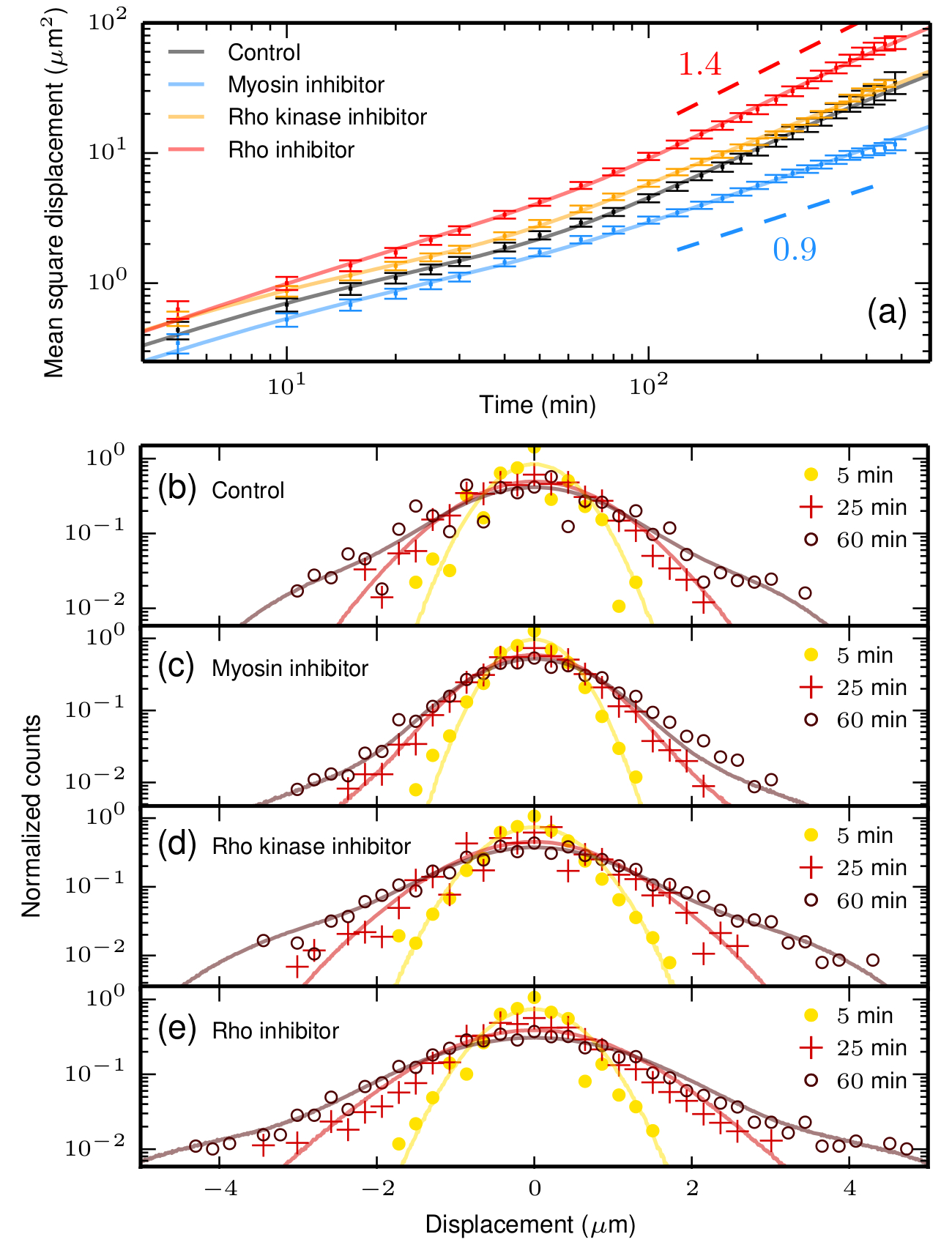}
\caption{\label{fig:dat}
	Statistics of vertex displacement.
	(a)~Mean square displacement as a function of time in four conditions: control (black), myosin inhibitor (blue), Rho kinase inhibitor (orange), and Rho inhibitor (red). The corresponding best fitting curves are in solid lines. The blue and red dashed lines report the large time behaviors.
	(b-e)~Distribution of displacement for the four conditions at three times: $5$ ($\bullet$), $25$ ($+$), and $60$~min ($\circ$). Exponential tails appear at long times as a consequence of directed motion events in  vertex dynamics. Results of simulated dynamics are in solid lines.
}
\end{figure}

We compute the projected one-dimensional mean square displacement (MSD) within the four different conditions [Fig.~\ref{fig:dat}(a)]. For each condition, the short time MSD exhibits a power-law behavior with exponent close to $0.7$ over about one decade. Interestingly, a subdiffusive behavior was reported for the dynamics of vertices in the endoplasmic reticulum~\cite{Lippincott2016}. The large time MSD depends on conditions and exhibits a behavior that, for simplicity, we have characterized by a power-law. The corresponding exponent is typically larger than $1$, except for myosin inhibitor where it is smaller. The crossover between the two regimes appears between $20$ and $60$~min. Fluctuations are reduced in the myosin inhibited case, which has the lowest MSD, and they are enhanced for the Rho inhibitor, where the long time MSD is the largest. We also explore the full statistics of vertex displacement by measuring the probability distribution function (PDF) for each condition, as shown in Figs.~\ref{fig:dat}(b-e). At short time the PDF is Gaussian, while it exhibits broader tails at large time. These tails reveal large displacements of the vertex, and were already observed for tracer particles in active gels~\cite{Toyota} and living cells~\cite{Park2010, Nir, Trepat:09, Bursac2005, Fodor}. They are more pronounced in the Rho inhibitor case, as a signature of larger fluctuations, possibly due to directed motion events.

\section*{Phenomenological model of vertex dynamics: transient confinements and large displacements}

To quantitatively discriminate between the effects of the different inhibitors, we analyze our measurements with a nonequilibrium model previously introduced to describe tracer fluctuations inside living cells~\cite{Fodor}. This model is not aimed at describing any specific process that produces the active fluctuations, it rather formulates a general framework that allows one to quantify non-equilibrium forces and fluctuations. We regard the vertex as a virtual particle which dynamics is prescribed by two coupled equations: (i)~an equilibrium diffusion of the vertex in a cage, modelled as an harmonic potential of stiffness $k$---the displacement is driven by a Gaussian white noise of variance $2 \gamma \kb T$ with a drag force of coefficient $\gamma$; (ii)~a non-Gaussian colored diffusion equation for the center of the cage, mimicking nonequilibrium activity as a run-and-tumble dynamics. Inspired by the large ballistic-like displacements that we observe in experimental trajectories, we model this active noise as a two-state Poisson process: the cage has a constant velocity $v$ in a random uniformly sampled two-dimensional direction during a random persistence time of average $\tau$, and it remains fixed during a random quiescence time of mean $\tau_0$. We understand the confinement as an elastic mechanical stress resulting from cells surrounding each vertex, and the nonequilibrium motion of the cage as an active stress. The effect of this active stress is to reorganize the structure of the monolayer, and therefore to spatially redistribute the elastic mechanical stress.

\begin{figure}[b]
\includegraphics[width=\columnwidth]{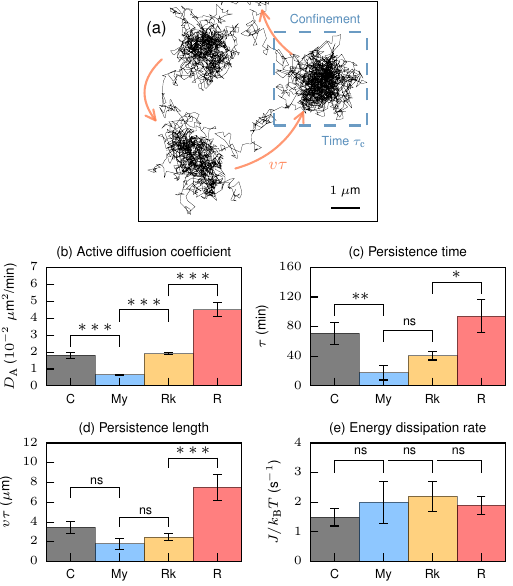}
\caption{\label{fig:traj}
	Active parameters of vertex fluctuations.
	(a)~Typical trajectory obtained from simulations of the vertex dynamics in control condition (scale bar $1$~$\mu$m). Isotropic ``blobs'' reveal equilibrium-like transient confinement during a typical time $\ts$ (dashed blue box), and large displacements of order $v\tau$ occur due to nonequilibrium activity (orange arrows).
	(b)~Best fit values of the active diffusion coefficient, (c)~the persistence time, (d)~the persistence length, and (e)~the energy dissipation rate.
	Statistical analysis with t-test:  ns (non significant) $p > 0.05$, * $p < 0.05$, ** $p < 0.01$, *** $p < 0.001$ (see the Supporting Material).
}
\end{figure}

In the absence of activity, this model predicts a short time diffusion, and then a large time plateau expressing the elastic confinement. Such dynamics is entirely under the control of equilibrium thermal fluctuations. In an active system, nonequilibrium processes enhance vertex displacement \textit{via} the cage motion, yielding a free diffusion of the vertex with coefficient $\Da=(v\tau)^2 /[2(\tau+\tau_0)]$. The large time dynamics is fully determined by the active parameters $\{v,\tau,\tau_0\}$, whereas thermal fluctuations control the short times \textit{via} $\{k,\gamma,T\}$. A sub-diffusive transient regime appears between the two diffusions, as a crossover towards a plateau, and a super-diffusive regime can also precede the large time diffusion, as a signature of the ballistic motion involved in the active noise. In such a case, thermal effects are negligible at times larger than $\ts=\sqrt{\tau\kb T/(k\Da)}$, a timescale quantifying the transition from the short time equilibrium-like dynamics to the large time active diffusion. Simulated trajectories exhibit clusters of similar size accounting for the transient confinement of the vertex. Occasionally large displacements of order $v\tau$ appear. The vertices do not only fluctuate around a local equilibrium position, they also undergo rapid directed jumps (compare Fig.~\ref{fig:exp}(e) and Fig.~\ref{fig:traj}(a)).

In contrast to previous works which describe the many-body dynamics of cells in the tissue in a more complete framework~\cite{Farhadifar:2007p4500, Kafer:2007p4501, Manning:15b}, we do not explicitly account for interactions between neighbouring vertices. As a result of such interactions, the cells experience an intermittent dynamics alternating between fluctuations of small amplitude and rapid large displacements. As in glassy systems, the jumps appear through collective rearranging regions~\cite{Bi2016}, thus contributing to the non-Gaussian fluctuations experienced by the vertices. Even though non-Gaussian fluctuations are present independently of topological transitions, the existence of a quantitative connexion between collective rearrangements and such transitions is still an open question~\cite{Bi2014}. Our approach consists in reducing the dynamics of a large number of interacting vertices into the dynamics of a single vertex embedded in an effective background that describes the mean-field effect of the surrounding system. Within our model, interactions are embodied by both the elastic confinement and the active source of fluctuations leading to large displacements. We argue below that our phenomenological approach is sufficient to capture the vertex dynamics, since it provides a framework to decipher the effects of the Rho pathway inhibitions on this dynamics.


\section*{Order relation in the active diffusion coefficient and the persistence time}

We fit the MSD data with our analytic prediction to estimate the parameters characterizing nonequilibrium activity. Our fits convincingly capture the transition from sub-diffusive to super-diffusive like behaviors [Fig.~\ref{fig:dat}(a)]. Within our model, these behaviors correspond to cross-over regimes between the short and large time diffusion. We extract from the best fits a single set of passive and active parameters for each condition. Our estimate for $\ts$ can be compared with timescales quantifying the transition from elastic to fluid-like behavior of the material (see Tab.~S1 in the Supporting Material), which is of the same order as the Maxwell time reported in three-dimensional cell agregates, \textit{i.e.} about $30$-$40$~min~\cite{Forgacs, Guevorkian, Marmottant, Stirbat}. We report clear quantitative variations of both the active diffusion coefficient $\Da$ and the persistence time $\tau$ for all conditions [Figs.~\ref{fig:traj}(b,c)]. This suggests that our model, based on separating purely active fluctuations from equilibrium thermal ones, is a reliable framework to capture the effects of our inhibitors on tissue dynamics. Note that passive parameters, such as the relaxation time scale $ \tau_\text{r} = \gamma / k $ reported in Tab.~S1 (see the Supporting Material), have also different values between the conditions. This reflects the effects of inhibitors on tissue mechanics, showing that inhibitors also affects the characteristics of passive fluctuations. In this respect, $\Da$ and $\tau$ are the parameters that characterize {\it only} the active contribution to vertex fluctuations, which is the main focus of our study.

Strikingly, $\Da$ and $\tau$ are larger for Rho inhibitor than for Rho kinase inhibitor, and than for direct myosin inhibition [Figs.~\ref{fig:traj}(b,c)]. The more upstream the inhibition along the pathway, the larger the amplitude of fluctuations and the more persistent the ensuing displacement. The myosin inhibition leads to the smallest $\Da$ and $\tau$ values, suggesting that the mesoscopic activity of vertices is strongly affected. The Rho kinase target directly activates the myosin, but it also inhibits the myosin light chain phosphatase (MLCP) which in turn inhibits the myosin. Therefore, the result of Rho kinase inhibition on myosin can not be anticipated \textit{a priori}. Our analysis shows that activity of vertices is less affected than for the myosin inhibitor case: the corresponding value of $D_\A$ for Rho kinase inhibitor is close to the one for control, which suggests a compensation between activation and de-activation of myosin.


\section*{Order relation confirmed by persistence lengths}

To gain further insight into the active component of the dynamics, we compare the displacement PDF extracted from the simulated trajectories of vertex dynamics with experimental distributions. The distribution at short time is Gaussian and entirely controlled by the passive parameters: the simulations with or without active fluctuations, where we use passive parameters estimated from fits of MSD data, give the same results at short times (see yellow curves in Figs.~\ref{fig:dat}(b-e)). Including the active component for the dynamics leaves us with one remaining free parameter: the average persistence length $v\tau$ of large displacements. The short time Gaussian remains unchanged, whereas exponential tails develop at large times in the simulated PDF. The tails are more pronounced as time increases, while the central Gaussian part barely changes. We adjust the $v\tau$ value by matching the tails appearing in numerical results and experimental data.

The simulated PDFs compare very well with experiments at large times, showing that our simulations reproduce the evolution of experimental distributions at all times [Figs.~\ref{fig:dat}(b-e)]. This is one of the main success of our analysis: the phenomenological model on which relies the quantification of internal activity is able to capture the strong non-Gaussian tails of the distribution with only one free parameter. This supports the underlying picture that vertex dynamics essentially alternates between transient confinement and directed motions. The order of magnitude of the extracted mean persistence length $v \tau$ is consistent with our measurements [Fig.~\ref{fig:traj}(d)]. We report again the same order relation within the Rho pathway, \textit{i.e.} an increase of $v\tau$ from myosin inhibitor to Rho kinase inhibitor, and from Rho kinase inhibitor to Rho inhibitor, as a signature of enhanced directed motion.


\section*{Dissipation is constant for all inhibitors}

A major asset of our model is that it allows us to predict the mean rate of energy dissipated by the vertex dynamics in its surrounding environment. It is defined as the difference between the power injected by the fluctuating thermal force and the one that the moving vertex dissipates \textit{via} the drag force: $J=\avg{\dot x(\g\dot x-\sqrt{2\gamma\kb T}\xi)}$, where $\dot x$ is the vertex velocity, and $\xi$ is a zero-mean Gaussian white noise~\cite{Sasa, Sekimoto}. It vanishes for systems in a thermodynamic equilibrium state. The active nonequilibrium fluctuations lead to a non zero dissipation rate: $J=k\Da/(1+\tau/\tau_\text{r})$~\cite{Fodor2}. This rate of dissipated energy reflects the excess power injected by nonequilibrium internal activity driving the large displacements of vertices.

When computing the dissipation rate in the four conditions, it appears as approximately constant [Fig.~\ref{fig:traj}(e)], in contrast with the order relation found for active parameters [Figs.~\ref{fig:traj}(b-d)]. This supports that the same amount of energy is dissipated by the vertex large displacements, though the features of such displacements differ between conditions. Given that $J$ depends both on parameters of active fluctuations $\{\Da, \tau\}$ and on parameters of passive mechanics $\{\gamma, k\}$, our result suggests that there may be an underlying coupling between mechanical properties of the tissue and its nonequilibrium fluctuations. A possible interpretation is that the nonequilibrium processes at the origin of active fluctuations, such as forces induced by myosin and by actin polymerization, might also affect the tissue mechanics in such a way that the dissipation rate remains unchanged over all the conditions. In that respect, we observe that the relaxation time $\tau_\text{r}$ is increased for Rho inhibitor case with respect to others (see Tab.~S1 in the Supporting Material), as also observed for the persistence time $\tau$ [Fig.~\ref{fig:traj}(c)].


\section*{DISCUSSION}

The parameters of vertex fluctuations reveal a correlation between the Rho pathway hierarchy and the junction active fluctuations: the higher up the Rho pathway is the inhibition, the lower is the effect on decreasing the fluctuations of the vertices. Such a relation highlights the usefulness of our methodology to probe quantitatively how signaling pathways control emergent fluctuations. In that respect, our approach bridges biochemical signaling pathways with mechanical fluctuations {\it in vivo}. It could be used to characterize quantitatively the effect on fluctuations of other inhibitions acting on different signaling pathways.

Our analysis is based on a phenomenological model which deliberately avoids a detailed description of the many processes operating at the subcellular scale. It is also distinct from other models which consider explicit interactions between neighbouring cells. In that respect, our results do not rely neither on the microscopic details of activity-induced forces nor on the form of interactions within the tissue. We rather postulate an effective vertex dynamics by explicitly distinguishing thermal and active fluctuations. As a result, any underlying mechanism which could rationalize the relation between pathway inhibition and mechanical fluctuations is out of the scope of our approach. Yet, the relation between biochemical signaling and active fluctuations that we quantified in tissues could help future studies to understand the precise relation between the subcellular cytoskeleton dynamics and the emergent vertex fluctuations. In that respect, a possible extension of the model could consist in considering more complex activation/inhibition in the Rho pathway and their connection with contractile forces in the cell~\cite{Besser2007, Bement2015}.

In order to test the robustness of our model, one could measure the response of vertices to an external perturbation~\cite{Harris2012}. The intracellular mechanics is viscoelastic in a large variety of living systems. Our phenomenological model has already been extended to account for a complex rheology~\cite{Ahmed2016}. It would be interesting to determine whether the relation between fluctuations and pathway inhibitions is similar when including viscoelastic effects. Moreover, applying an external potential to confine a vertex, one could extract work from the vertex fluctuations by varying in time the potential parameters. Our framework allows one to predict the details of the extracted work as a function of the active fluctuation characteristics~\cite{Fodor2}. Confronting such predictions with experimental results would provide another test for the validity of our approach.

The response in living systems is not related to the spontaneous fluctuations, in contrast to equilibrium where such a relation is given by the fluctuation-dissipation theorem (FDT). Therefore, by comparing the response with spontaneous fluctuations of tissues, one could quantify the departure from the FDT, namely the deviation of the dynamics from equilibrium. Moreover, recent progress in nonequilibrium statistical mechanics have shed light on the direct relation between the rate of energy dissipated by the internal nonequilibrium processes and the violation of the FDT~\cite{HarSas}. This has already led to access to the dissipation rate in some biological contexts~\cite{Toyabe2010, Ahmed2016}. In that respect, comparing response and fluctuations will allow one to propose an alternative quantification of dissipation rates in epithelial tissues.

The Rho pathway is conserved across species, suggesting that regulations of activity in epithelial monolayers may share common pathways in a large variety of tissues and organisms~\cite{Hall2002}. Our approach, based on a quantitative analysis of vertex fluctuations, could serve as a novel framework to decipher the complex regulation of spatial fluctuations by the Rho pathway in other model systems. In that respect, it could be used to analyze vertex fluctuations in developing embryos, such as in {\it Drosophila} or in {\it C. elegans}, where internal reorganizations are driven by spontaneous topological transitions~\cite{Rauzi:08, Guillot:2013}.

\section*{Author contributions}

V.M., J.C., R.T., and D.R. designed and performed the experiments. {\'E}.F., N.S.G., P.V., and F.v.W. designed the model. All authors contributed to analyzing the data and writing the manuscript.

\section*{Acknowledgements}
We thank F. Graner for helpful discussions. We acknowledge J. W. Nelson for sending the MDCK cell lines. We also thank N. Maggartou for extraction of data and lab. members for discussions. This work is supported by CNRS, FRC and University of Strasbourg. N.S.G. is the incumbent of the Lee and William Abramowitz Professorial Chair of Biophysics.


\section*{SUPPORTING MATERIAL}

An online supplement to this article can be found by visiting BJ Online at http://www.biophysj.org.


\bibliography{Epithelial_ref}

\end{document}